# Absolutely Secure Communications by Johnson-like Noise and Kirchhoff's Laws

Laszlo B. Kish *

**Abstract:** We survey the most important results and some recent developments about the secure key exchange protocol where the security is based on the Second Law of Thermodynamics and the robustness of classical physical information. We conclude that a classical physical system offers a higher level of control and security during the communication. We also mention some recent attempts inspired by this communicator to create other systems where Alice and Bob do not form an organic single system and/or the Second Law is irrelevant. It seems philosophically that they cannot be unconditionally secure, however it is yet an open question how to crack them; how can they be best used for conditionally secure communications, and what are the practical implications.

## 1. Introduction

Very recently, it has been shown that electrical random noises can be utilized for communication and computing as information carrier with various peculiar properties. Examples are Kirchhoff-Law-Johnson-(like)-Noise (KLJN) based secure communication [1-3], Zero-Signal-Power (stealth) Communication [4], and Noise-based logic (continuum [5,6] or spike [7] noise based - brain-mimic). One of the interesting results of these studies is that these noise-based schemes, which are classical physical systems, look competitive alternatives of quantum informatics: both quantum communication [1-3] and quantum computing [6]. Moreover [7], they serve a potential explanation to explain the brain logic and to mimic this brain logic scheme by electrical circuits.

## 2. Secure key exchange utilizing the laws of physics

In today's software-based secure communications (tools we use when connect our bank via the internet), before the secure data exchange can start, the two communicators (Alice and Bob) must generate and share a joint secret (secure) encryption key through the communication channel while the eavesdropper (Eve) is supposedly monitoring the related data (Fig. 1). This is a mathematically impossible task with current software methods thus they are only "computationally safe" that is Eve can decode the data but it takes too long time. Thus, if Eve had a genuine powerful algorithm or a sufficiently fast computer with standard algorithms, she could extract the secure key and decrypt the communicated data with a reasonable speed. Because new algorithms and computing solutions are continuously researched, today's software-based secure communication is a potential time bomb.

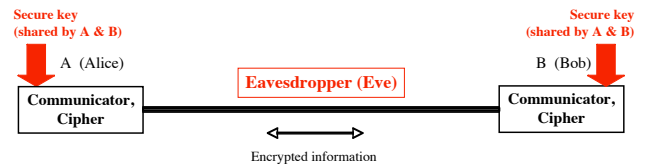

**Figure 1.** The two communicators (Alice and Bob) must generate and share a joint secure key through the communication channel while the eavesdropper (Eve) is monitoring the related data. This is an impossible task with software methods thus currently used software methods are only "computationally safe", which means they are potential time bombs.

Quantum key distribution (see [8-10] and references therein), due to Stephen Wiesner (1970's); Charles H. Bennett and Gilles Brassard (1984); and Artur Ekert (1990), has offered a solution that is claimed to be unconditionally secure. The information bits are carried by single photons (Fig. 2). Here the *no-cloning-theorem* of quantum physics is the theoretical foundation of security. It means that a single photon cannot be copied without noise (error). If Eve captures and measures the photon, it gets destroyed and she must regenerate and reinject it into the channel otherwise this bit will be considered invalid by Alice and Bob. However, due to the no-cloning rule, while Eve is doing that, she introduces noise and the error rate in the channel will become greater than without eavesdropping. Therefore, by evaluating the error statistics, after analyzing a number of transmitted bits and their errors, Alice and Bob will discover the eavesdropping with a certain probability. However, no quantum communicator is secure against the advanced type of the man-in-the-middle-attack, where Eve breaks the channel and installs two quantum communicators in the line. With one of them she will communicate with Alice and pretends that she is Bob and with the other one she will communicate with Bob and pretends that she is Alice. This is one example, where the secure wire communicator described in the next section is superior to quantum encryption.

Many quantum communicators have been reportedly built, up to the range of a ≈200 km model line, (Toshiba and NTT) where the bit exchange rate is less than 0.25 bit/second [10]. Most of them are working through optical fibers and some of the most advanced and secure ones are able to communicate via air. However, the experimental testing of different breaking ideas can be more expensive than to build the communicators themselves! Therefore, because this technology is extremely expensive, today's quantum security against the various breaking attempts is mostly theoretical, with a vast amount of theoretical-only papers about proposed breaking methods. It means that, say, only about 10% of the necessary experimental work is done for the existing prototypes (because there are many theoretical ways to break into a quantum channel) and, say, the 90%, which is testing and enhancements based on the test, must still be done before these quantum devices can be marketable on a large scale.

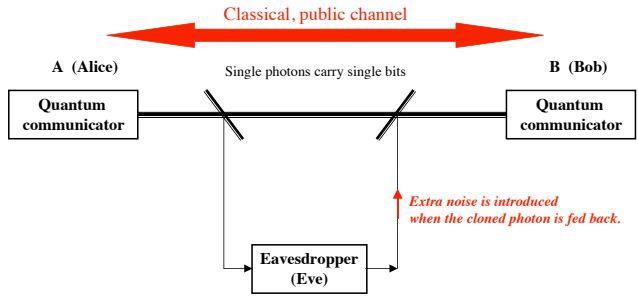

**Figure 2.** Generic quantum communication arrangement. To detect the eavesdropper, a statistics of bit errors must be built. That requires a sufficiently large number of bits. The communication of just a few bits is not secure.

## 3. The secure communicator based on classical thermodynamics

Recently, an unconditionally secure classical-physical communication scheme, the Kirchhoff-Law-Johnson-(like)-Noise (KLJN) communicator has been proposed [1-3, 11], which is a statistical-physical competitor of quantum communicators. The security against *passive attacks* (passive voltage and/or current measurements) is based on the Second Law of Thermodynamics: that is the impossibility of building a perpetual motion machine of the second kind. Its security against *active attacks* (injecting or extracting current from the channel) is based on the robustness of classical physical information implying that the current and voltage data can be continuously monitored in the line. The KLJN system contains two identical pairs of resistors (Fig. 3). The logic-low (L) and logic-high (H) resistors, $R_0$ and $R_1$, are randomly selected at the beginning of each clock period and are driven by their own Johnson-noise (thermal noise) voltages or alternatively by the electronically enhanced versions (Johnson-like noise) with a publicly agreed common, nominal temperature. The practical realizations contain more elements, such as filters, amplitude control units, etc. A secure key bit is generated and exchanged when the resistor values at the two ends differ.

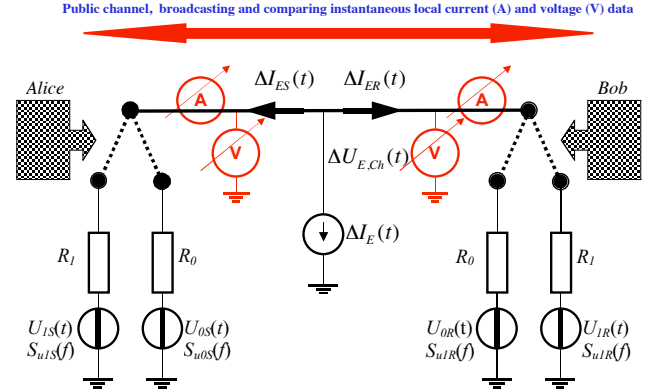

**Figure 3.** The fully protected KLJN system. To detect the invasive eavesdropper (represented for example by the current generator at the middle), the instantaneous current and voltage data measured at the two ends are broadcasted and compared. The eavesdropping is detected immediately, within much shorter period than the time needed to transfer a single bit. Statistics of bit errors is not needed. The communication of even a single bit is secure. Low-pass line filters are necessary to protect against out-of-alarm-frequency-band breaking attempts. False alarms would occur due to any wave effect (transient or propagation effects), illegal frequency components or external disturbance of the current-voltage-balance in the wire.

The role of the Johnson (-like) noise is the determination of the total resistance in the loop without serving information to Eve about the actual location of $R_0$ and $R_1$. When the total loop resistance gets known (this information is available to the public, too) and when it is the sum of $R_0$ and $R_1$, Alice and Bob can calculate the resistance value at the other side since they know their own resistance values.

The KLJN cypher is naturally protected against the man-in-the-middle attack [11], see Fig. 4, and, in the same way, the active eavesdropping is detected immediately [1,11], within much shorter period than the time needed to transfer a single bit. Statistics of bit errors is not needed. The communication of even a single bit is secure.

It is important to note that both the quantum and classical claims about unconditional security outlined above are about the idealized systems (mathematical model level). In practical applications, no physical system is ideal and there are parasite effects and elements. Therefore, in practical applications, neither the quantum nor the KLJN systems are totally secure. However, knowing their mathematical model, their security and other performance can be designed depending on physical and financial limits. The ultimate test of security must be experimental: before marketing a secure communicator must be tested by all the known breaking methods.

In [12] it was pointed that, due to the continuous monitoring and comparison of the voltage and current data at the two ends of the channel, Alice and Bob always exactly know Eve's information provided Eve is making the best possible guessing. Then Alive and Bob exactly knows if Eve is right or wrong when she estimates a key bit. This feature is another unique characteristics of the KLJN system among secure communicators.

In [13], a KLJN network scheme was developed which is able to generate and share keys with high speed over a chain network of KLJN units instead of the usual point-to-point connected systems.

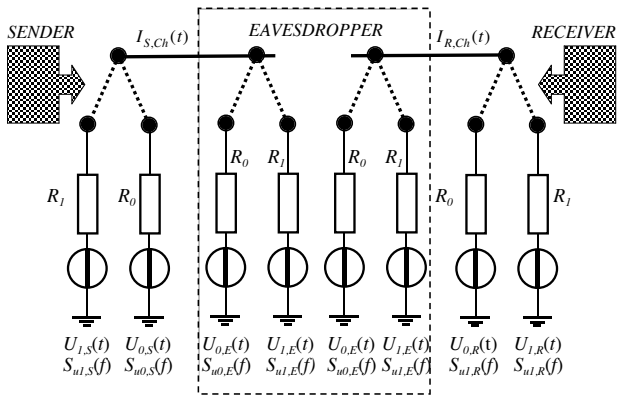

**Figure 4.** An example for the man-at-the-middle-attack when Eve uses resistors with the same values and noise voltage generators with the same parameters as those of the sender and the receiver [11]. In the two separate loops the current noises are totally independent thus the attack is immediately discovered within a fraction of the clock duration. Zero bit can be extracted and the alarm goes on. The man-in-the-middle-attack is one of the weakest type of attacks against the KLJN cipher. This property is unique among known communicators.

There have been several attempts of breaking into the KLJN cypher, see: [14] with response [15]; [16] with response [17]; and [18] with response [12]. However neither of them, not even the correct proposals, could extract any information from the idealized system. [14] and [16] utilized non-idealities to create a small information leak, which can however be arbitrarily reduced (depending on resources) by proper design approaching the idealistic situation. The remaining information leak can be removed by privacy amplifiers, which are software-based tools used also by quantum communicators for similar purpose. Out of these, the unique feature mentioned above that Alice and Bob always knows when Eve is right or wrong, can also be utilized for defense [12].

Finally, the cracking claims in [18] were based on strongly unphysical assumptions (see the proof in [12]), such as a cable impedance corresponding to cable diameters greater by 28000 times than the size of the known universe [12]. Moreover, in [18], a circulator-based "imitation" of the KLJN system was introduced with the claim that it is more advantageous than the KLJN cypher. However, in [12] it was shown that the circulator-based system is vulnerable against the man-in-the-middle-attack which means it is not a competitor of the original KLJN system.

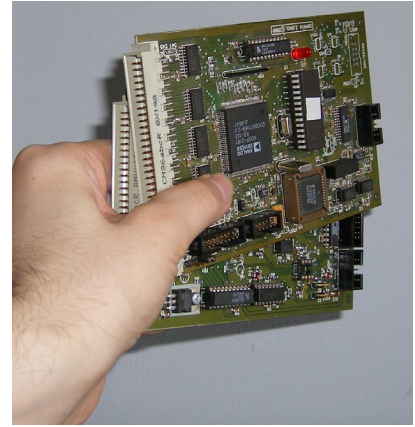

**Figure 5.** The KLJN wire communicator network element (communicator pair) tested for ranges of 2-2000km. Its fidelity is 99.98% and it is protected against all known types of attack.

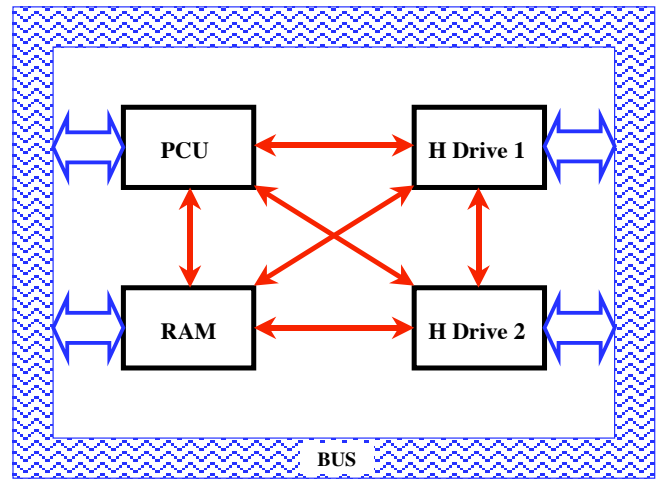

**Figure 6.** Example for securing a subsystem of a PCU, a RAM and two hard drives. Solid arrows: *KLJN* connections; Block arrows: classical data bus connections. Each unit has 3 *KLJN* communicators integrated on their chip. The required number of KLJN units/chip is $N$-1 when $N$ chip gets connected securely.

Due to the young age of the KLJN idea, so far, only one system has been built and tested [19] up to 2000 km range with model line, see (Fig. 5). It was experimentally verified against all the proposed invasive attack types and, in all cases, the invasive eavesdropping was discovered during the communication of a single bit. It had 99.98% fidelity, and 0.19% information leak of the raw bits during passive eavesdropping. The speed of the key exchange at 200 km range was 1 bit/second [19], which is 4 times faster than the Toshiba-NTT quantum key exchange system, at the same distance [10]. Its price was a few hundred dollars and, in an integrated form, its fabrication price will be similar to that of an Eternet card in a PC.

It is an important practical aspect that the KLJN system can be integrated on computer chips and secure communication between units in a computer or machine can be established; see Fig. 6 [20]. This property is also a unique feature of the KLN system.

## 4. Can we avoid the single physical system and the Second Law of Thermodynamics?

Recently, an interesting method was shown by Pao-Lo Liu [21] where instead of using a physical system simply random numbers were sent an reflected between Alice and Bob while new random numbers (noise) were added to the signal at each step. The protocol has been inspired by the KLJN system and, at the first look, it may seem to be secure. However, the system is neither physical nor it is related to the Second Law of Thermodynamics. Moreover, it must philosophically be obvious that such a system, where Alice and Bob are physically separated, can hardly be a good model of the KLJN system, where properly the limited bandwidth connecting the resistances guarantees that Alice and Bob form a single physical system that they can manipulate simultaneously. In the protocol described in [21] Eve can freely observe all the number values sent back and forward and she has significant information for cracking.

Indeed, on the contrary of its very young age, this protocol was already cracked in its most secure state (the steady bit state) [22] for a large class of noises that can be transformed to white noise in the active frequency range by filtering. Even though, Alice and Bob can choose an idealistically band-limited noise, during changing the bit and the related transients, the noise will lose its band-limited nature and the system becomes vulnerable. Even though this effect can be reduced by line filters, work in progress indicates that similar methods can be developed for arbitrary types of noises.

Finally, we mention an interesting idea [23,24] of Scheuer and coworkers which was also inspired by the KLJN cypher. The oscillation frequency in a long laser is varied by Alice and Bob by selecting and placing filters at the two ends of the laser. Similarly to the resistances in KLJN, they have identical pairs of optical filters with different frequency characteristics within a pair. This system is physical however the Second Law of Thermodynamics is not relevant. Furthermore, the photons can be detected and measured in both directions because this system is not an integral one either. The same philosophy which lead to crack the Liu cypher suggests that this system may also be non-secure at the unconditional level. However, how to crack it is still an open question.

The final question about these new key exchange protocols is the following: Though they can probably be cracked even at the idealized situation, how can we utilize their limited security at practical applications. Maybe, by using a high-security seed-key obtained from the KLJN cypher they can run for a longer period without the need of refreshing that seed?